\documentstyle[11pt,epsf]{article}
\newcommand{\tket}[1]{\left|\left|\left|#1\right\rangle\right.\right.}
\newcommand{\tbra}[1]{\left.\left.\left\langle#1\right|\right|\right|}
\begin{document}
\begin{titlepage}

\newcommand\letterhead {%
\hfill\parbox{8cm}{    \large \it UNIVERSITY of PENNSYLVANIA\\
\large Department of Physics\\
David Rittenhouse Laboratory\\
Philadelphia PA 19104--6396}\\[0.5cm]
{\bf PREPRINT UPR--0126MT}\\ June 1993}
\noindent\letterhead\par
\vspace{2 cm}
\noindent
\begin{center}
{\Large \bf From Skyrmions to $NN$ phaseshifts}\\
\end{center}
\begin{center}
{\bf \large
Niels R. Walet}
\end{center}
\begin{center}
\vfill
{\em Submitted to Physical Review C}
\end{center}
\end{titlepage}

\title{From Skyrmions to $NN$ phaseshifts}
\author{Niels R. Walet\thanks{electronic address:
walet@walet.physics.upenn.edu} $^,$\thanks{address after Sept.~1, 1993:
Instit\"ut f\"ur theoretische Physik III, Universit\"at Erlangen-N\"urnberg,
D-91058 Erlangen, Germany}\\
Department of Physics, University of Pennsylvania,
Philadelphia PA 19104-6396
}
\maketitle

\begin{abstract}
We study the $NN$ phaseshifts in a Hamiltonian obtained from
quantization of the collective modes in the Skyrme model.
We show that a combination of an adiabatic and diabatic
approximation gives a good $NN$ force, with sufficient attraction
to produce a bound deuteron. The description of the repulsive core
appears to be the main cause for the remaining discrepancies
between the Skyrme model force and phenomenology.
Finally we discuss the possibility of finding non-strange
dibaryon resonances in the $J^\pi=3^+$ channel.
\end{abstract}

\section{Introduction}

In this paper we continue the study of the derivation of a nucleon-nucleon
force from the Skyrme model. In previous work \cite{WaletAH92,WaletAmado93}
we have shown
that the Skyrme model gives a good qualitative description of the intermediate
and long range nuclear force. Here we  study
the phaseshifts obtained in the model. This requires new information about
the repulsive core of the potential. Many more details of the problem,
as well as some  alternative approaches are reviewed in
Refs.~\cite{OkaHosaka92,WalhoutWambach92,NymanRiska90}.

In our previous work we have discussed all the steps that are  necessary
to obtain a good result for the $NN$  force.
The starting assumption is that the Skyrme model \cite{Skyrme62}, a non-linear
field theory of interacting pions, is related to  QCD
in the limit of a large number of colors ($N_c$). Even though this
connection is uncertain, one can take the point of view
that in the long wavelength limit only the lightest degree of freedom,
the pion, plays a role. Thus one expects the model to be good for
phenomena where only distances larger than the wavelength of the
$\rho$ meson, the next lightest particle, play a role.

An interesting property of the Skyrme model is that there exists
a topologically conserved current, usually associated with
the baryon number $B$. The $B=1$ solution, the Skyrmion or hedgehog,
has been used to describe properties of hadrons to better than
30 \% $\approx 1/N_c$, giving numerical support to the model.
Once we take the Skyrme model to be a valid starting point
for the discussion of single baryons, the next step is to
study the interaction of two baryons. To this end one studies
a collective-coordinate manifold of $B=2$ Skyrmion states,
which is rich enough to describe   the dynamics of $NN$ states.
Initial studies of the $B=2$ system were based on the so-called product Ansatz,
which is still used as a paradigm for understanding the structure
of the manifold (see the discussion in Ref.~\cite{AtiyahManton93}).
This Ansatz describes the
superposition of two Skyrmions without any distortion.
The advantage of this Ansatz is that it allows us to construct all
states on the collective manifold in  terms of a few
invariant combination of six parameters. This in turn can be
used to determine the collective potential and kinetic energy
on all points of the manifold.
To understand why the product Ansatz is not good enough for
a realistic calculation of the $NN$ force,
 consider the potential energy in a little more detail.
It is calculated as the  difference between the energy of two interacting
Skyrmions and twice the energy of a single Skyrmion. This number is
typically
of the order of tens of MeV's, but is obtained from the difference of two
numbers of the order of $2~\rm GeV$. This is obviously  sensitive
to relatively minor flaws in the Ansatz.
The product Ansatz has several flaws, that make it unlikely that
it can lead to a correct description of the $B=2$ system.
These are first of all that the product Ansatz is
not symmetric under interchange of the two hedgehogs,
secondly that it  can not describe the cylindrically symmetric
ground state of the $B=2$ manifold, the donut
\cite{KopeliovicStern87,Verbaarschot87,Manton88a,BraattenCarson88}.

In order to find improved solutions one can try to find better solutions
on part of the collective manifold (the calculations are too
demanding to solve for every state).
To that end one picks out those sets of solutions that
have a definite reflection symmetry (see, e.g., Ref.~\cite{VWWW87}).
The Skyrmion configurations are now obtained numerically
by imposition of a constraint on the separation of
the two solutions, which leads
to distorted states, finally culminating in a state of toroidal
symmetry, the donut.
This method is based on the assumption that these solutions are
stationary against deformations that break the reflection symmetry.
{}From preliminary results of a calculation performing a local
RPA on the Atiyah-Manton Ansatz, this assumption appears to be largely
correct \cite{WaletRPA93}.
As stated above the disadvantage
of this  approach is that one  obtains only partial  information
about the collective Hamiltonian. Fortunately the information one
can extract coincides with the dominant part of the interaction
as calculated in the product Ansatz.

After performing all this analysis, we still end up with a classical
collective  Hamiltonian
that describes the large $N_c$ limit, where the nucleon and its excited
states are degenerate. Since we believe that the quantum world has
$N_c=3$ we cannot requantize the collective Hamiltonian as it stands.
Instead we requantize the Hamiltonian in a space of nucleons and $\Delta$'s.
We use an algebraic technique to perform this reduction
\cite{AmadoBO86,OBBA87}. This
leads to a coupled channels problem, and not yet to a $NN$ force. In
a final step we invoked the Born-Oppenheimer
approximation   in our previous work to
obtain the $NN$ adiabatic nuclear force, which was shown to
be similar to a phenomenological force, such as the Reid soft-core
potential \cite{Reidpot}.

In this paper we shall study the phaseshifts calculated
in the full ($N-\Delta$) case and for the adiabatic reduction.
This will give us information about the adiabatic reduction, as
well as about the quality of the intermediate coupled channels problem.
We will
have to go slightly beyond the adiabatic approximation in order to
obtain information about the short-range repulsive core.
This last part of the potential must be included in a
calculation of $NN$ phaseshifts.

Tes paper is organized as follows. In Sec.~\ref{sec:NN} we study
our model for the collective Hamiltonian and the reduction to
a $NN$ force. In the next section, Sec.~D (\ref{sec:phases}),
 we discuss
the phaseshift calculation for the $^1S_0$ and $^3S_1-$$^3D_1$ channels.
Some technical details about the calculations  can be found in the appendix.
In Sec.~\ref{sec:Delta}
we discuss the possibility of finding resonances near $\Delta$ production
thresholds. Finally, in Sec.~\ref{sec:conc}, we draw some conclusions.

\section{The model and the $NN$ reduction\label{sec:NN}}
\subsection{The Skyrme model}
The Skyrme model \cite{Skyrme62,ZahedBrown86,Liu87}
is a non-linear field theory that can be realized in
terms of an SU(2)-valued matrix field $U$, with Lagrangian density
\begin{equation}
{\cal L} = \frac{f_\pi^2}{4} {\rm Tr}[
\partial_\mu U(x) \partial^\mu U^\dagger(x)]
+ \frac{1}{32g^2}{\rm Tr}[ U^\dagger\partial_\mu U, U^\dagger\partial_\nu U]^2
+\frac{m_\pi^2}{4}[{\rm Tr}(U+U^\dagger)-4].
\end{equation}
The model is covariant, as well as invariant under global
SU(2)-rotations that are identified with the isospin symmetry. As was
discovered by Skyrme the model has a topologically conserved quantum number,
which is identified as the baryon number $B$.
The $U$ field is interpreted as a combination of a scalar $\sigma$ field
and an isovector pion  field, $U=(\sigma +
i\vec{\tau} \cdot \vec{\pi})/f_\pi$.
The $\sigma$ field is not an independent physical field due to the unitarity
constraint on $U$.

The standard time-independent solution to the classical field equations
for $B=1$ is the defensive hedgehog, where the pion field points radially
outward,
\begin{equation}
U_1(\vec{r}) = \exp( i\vec{\tau} \cdot \hat{r} f(r) ).
\end{equation}
The baryon number of this state is given by $B=(f(0)-f(\infty))/\pi=1$.
This solution breaks translational invariance, as well as the $O(4)$
spin-isospin symmetry. If we perform a global SU(2) isorotation on the state,
\begin{equation}
U_1(\vec{r}|A) = A^\dagger U_1(\vec{r}) A,
\end{equation}
we obtain a state of the same energy.

In the $B=2$ system we will frequently use the product Ansatz, which
is also used as a model to understand the more complicated numerical solutions.
 This
Ansatz makes use of the fact that the product of two $B=1$ solutions
has baryon number two. The most general Ansatz we can  construct
from  two hedgehogs consists of the product of two separated and
rotated hedgehogs,
\begin{eqnarray}
U_2(r|\vec{R} AB)&=& A^\dagger U_1(\vec{r}-\vec{R}/2) A
                     B^\dagger U_1(\vec{r}+\vec{R}/2) B
\nonumber\\
                 &=& U_2(r|\vec{R} CD)
\nonumber\\
                 &=& D^\dagger C^{1/2\dagger} U_1(\vec{r}-\vec{R}/2) C
                     U_1(\vec{r}+\vec{R}/2) C^{1/2\dagger} D.
\label{eq:PA}
\end{eqnarray}
In the last line of (\ref{eq:PA}) we have introduced the matrix $D$,
that describes the rigid isorotation of the whole system,
as well as a relative isorotation $C$. When $R$ is very large changing $C$ or
$D$ does not change the energy of the solution. For smaller $R$, $D$
still generates
a zero-mode (corresponding to broken isospin symmetry), but the energy will
depend on $C$. Again, the energy is also invariant under spatial rotation,
due to the conservation of angular momentum $\vec{J}=\vec{L}+\vec{S}$.

\subsection{Modelling the force}

As discussed in our previous work, the introduction of collective coordinates
leads to an effective Hamiltonian for the relative motion and relative
orientation (we shall use the word relative orientation as a synonym
for the relative isorotation).
In the calculations of Refs.~\cite{WalhoutWambach91,HosakaOA91} only a small
part of the collective surface was calculated: Using reflection symmetries
one studies three one-dimensional ``lines'' on the collective manifold
where the radial coordinate $R$ changes, but the orientation (analogous
to the matrix $C$ in the product Ansatz (\ref{eq:PA})) remains fixed.
This allows one to extract limited information about the collective
Hamiltonian, through an  expansion in invariants to first order.
Fortunately the product Ansatz, where the complete expansion can
be calculated \cite{OBBA87}, allows one to show that
higher order terms are small.

The resulting classical Hamiltonian, obtained after making some crude
approximations for the kinetic terms, still treats radial motion
on the same footing as isorotational motion, which separates the
nucleon from the $\Delta$ and higher $I=J$ resonances.
As is well known the appearance of unphysical resonances with higher $I=J$
is due to the close relation of the Skyrme model with the string
theory that describes the large $N_c$ limit of QCD.
In order to make
progress towards the physical world, we need to re-impose the restriction
$N_c=3$. The way we proceed (see \cite{WalhoutWambach92} for an alternative
but equivalent approach) is through a requantization of the spin-isospin
degrees of freedom of the individual Skyrmions using an $SU(4)$ interacting
boson model for those degrees of freedom. The important step
\cite{AmadoBO86,OBBA87} is to identify the
boson number in these models with the number of colors. The representation
space then consist of twenty states: the nucleon and $\Delta$ with
all possible spin-isospin projections.

The requantized Hamiltonian
can be given in the following form \cite{WaletAmado93},
\begin{equation}
H = E_0+\frac{\hbar^2}{2M}(R^{-2}\partial_R R^2 \partial_R +L^2/R^2)+
\frac{1}{4\Lambda}(I_1^2+I_2^2+S_1^2+S_2^2)+
 v_1(R) + v_2(R) W + v_3(R)Z(\hat R).
\label{eq:Ham}
\end{equation}
In the potential energy (the last three terms in Eq.~(\ref{eq:Ham}))
the functions $v_i$ are taken from the calculation of the interaction
energy of two Skyrmions.  The operators $W$ and
$Z$ are similar to the spin-spin and tensor operators:
\begin{eqnarray}
W & = & T^{\alpha}_{pi} T^{\beta}_{pi}/N_{\rm C}^2,\nonumber\\
Z & = & T^{\alpha}_{pi} T^{\beta}_{pj}
[3\hat{R}_i \hat{R}_j-\delta_{ij}]/N_{\rm C}^2.
\end{eqnarray}
Here $\alpha$ and $\beta$ label
two different sets of bosons, used to realize the $u(4)$ algebras,
and $T$ is a one-body operator with spin and isospin $1$.
In order to simplify the algebra, we introduce a spherical
tensor notation, which allows for the use of the standard Racah
calculus. We then need the
 spin-isospin doubly reduced matrix elements of (the spherical tensor
form of ) $T$. These are given by
\begin{eqnarray}
\tbra{N} T  \tket{N}
&=& -10, \\
\tbra{\Delta} T  \tket{\Delta}
&=& -20, \\
\tbra{\Delta} T \tket{N}
&=& -8\sqrt{2}.
\end{eqnarray}

In the kinetic energy we have made the {\em approximation} that the mass is
a constant, equal to twice the relevant reduced mass (so $M$ still is an
operator in isospace). We assume that both the mass $M$ and
the moment of inertia
$\Lambda$ are independent of $R$, and
 that $\Lambda$ is independent of the orientation --
certainly a gross simplification.
Some preliminary work has been done on lifting those approximations
\cite{ShaoWA93},
but not enough is known at this instant to make a better approximation.

As in our previous work we  extracted the functions $v_i$ from the work
in Ref. \cite{WalhoutWambach92}.
Instead of using a table of numbers, we chose to make
a fit to the results. One reason is that we shall need to push the
calculation to short distances. We choose a form of powers of $r$
times exponentials of one- and two-pion range:
\begin{eqnarray}
v_i(r)& = &(a^1_i + a^2_i/r) \exp(-rm_\pi)
	  +(b^1_i r + b^2_i) \exp(-r2m_\pi),
\;\;i=2,3,\\
v_1(r)& = &(b^1_1r+b^2_1 + b^3_1/r + b^4_1/r^2) \exp(-r2m_\pi).
\label{eq:vfit}
\end{eqnarray}
The relevant parameters are listed in Table \ref{tab:vfit}.
\begin{table}
\caption{Parameters in the fit to the potential $v_i$, as evaluated
by \protect{\cite{WalhoutWambach92}}.
The unit for $r$ is fm, all energies are given in MeV. \label{tab:vfit}}
\begin{center}
\begin{tabular}{c|r|r|r}
$i$ & 1 & 2 & 3 \\
\hline
$a^1_i$ & --        & 1091.7 & 1618.7\\
$a^2_i$ & --        & -3156.4& -4687.1\\
$b^1_i$ & -5810.5   & 8964.3 &13268.7\\
$b^2_i$ &  22937.2  & 117.47 &478.23\\
$b^3_i$ &  -33370.8 & --     & --\\
$b^4_i$ & 17532.7   & --     & --\\
\end{tabular}
\end{center}
\end{table}

Note that the repulsive core is not very well determined from the standard
calculations \cite{WalhoutWambach92}. This is due in part to the uncertainty
in the definition of the coordinate $R$ for small separations
($R<1.3~\rm fm$).

\subsection{Calculation of  the $NN$ potential}
The calculation of an $NN$ potential from Eq.~(\ref{eq:Ham}) can be split
into two parts. One is the adiabatic calculation, valid when the $NN$
force is weak compared to the total energy  of the system
(from a numerical calculation one finds that, roughly, $R > 1~ \rm fm$).
As discussed in
Ref.~\cite{WaletAmado93},
this is based on the separation of energy scales:
For large $N_{\rm C}$, one can distinguish two
energy scales or reciprocally two time scales. The  slower time scale is
associated with the motion in the collective manifold, i.e., $R$ and the
orientation. The other time scale corresponds to the almost
instantaneous response of the
pion field to changes in $R$
and the relative isospin orientation.
For large $N_{\rm C}$ we cannot separate the time scales
for the two sets
of adiabatic modes, as can be seen in the highly correlated doughnut.
In the donut state orbital modes and isorotational modes are intrinsically
linked, leading to the interesting structure of this state.

For $N_{\rm C}$ equal to three the situation changes. The $R$ motion is
typically much slower than the rotational motion which leads to the
separation of the nucleon and $\Delta$ states. We thus have three energy
scales,
that of the pion field, of the $N-\Delta$ separation and of the $R$ motion.
We can now calculate a Born-Oppenheimer potential for the
$R$-motion, which constitutes the slowest degree of freedom
\cite{WaletAH92,WaletAmado93}.

This adiabatic -- Born-Oppenheimer --  approximation fails at shorter
distances ($R < 1~ \rm fm$), where the $R$ dependent potential
becomes strongly repulsive. Numerical experimentation showed that
for the calculation of the phaseshifts a {\em diabatic} prescription
seems to work well.
The adiabatic potential energy curves exhibit (relatively) narrow
avoided crossing in the interior region. If we follow
the adiabatic curve in the interior region,
we obtain a potential of rather poor quality.
If we follow the unmixed curve through the crossing, as if no crossing
took place, we obtain much better results. This corresponds to
a diabatic approximation, where we replace  avoided crossings by
real crossings.

\section{Calculation of phaseshifts \label{sec:phases}}
\begin{figure}
\epsfysize=8cm
\centerline{\epsffile{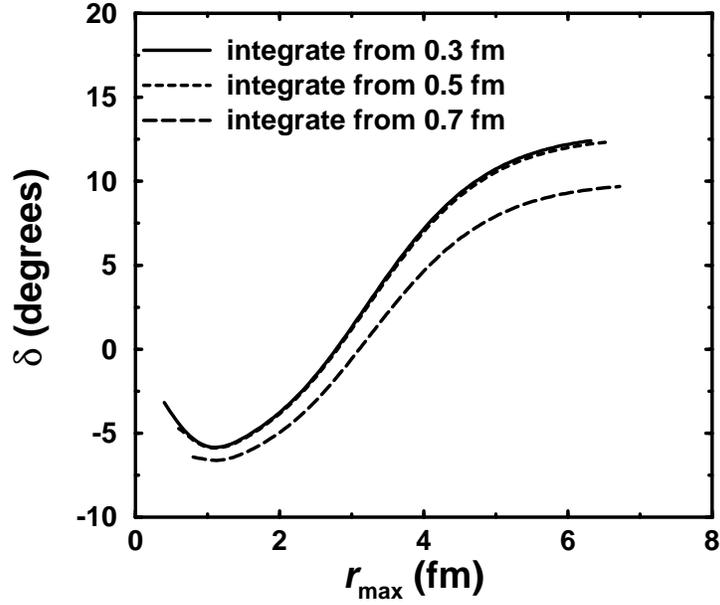}}
\caption{The phaseshifts in the $^1S_0$ channel ($E=1~\rm MeV$) as a
function of the matching radius where we match to the Bessel function
solutions of the free problem.
We compare calculations for three different hard core
radii.
\label{fig:match}}
\end{figure}

We calculate the phaseshifts by integrating the Schr\"odinger
equation from $r=r_{\rm min}$
where we impose hard-core boundary conditions (the forces are
so repulsive at short distances, that this is no limitation),
to $r=r_{\rm max}$
where we match to spherical Bessel functions describing the asymptotic
behavior of the wavefunctions (this corresponds to the approximation that
the potential vanishes beyond this point, but keeps all the slowly decaying
centrifugal forces.) As discussed in the appendix
this leads directly to an evaluation for the $S$ matrix,
which in its turn leads to the phaseshifts. Since we need to impose
hard core boundary conditions at $r=r_{\rm min}$, this point had
better be inside the repulsive core of the potential. The choice
of this interior radius is studied
in fig.~\ref{fig:match}, where we show the calculated phaseshift
in the $^1S_0$ channel for center-of-mass energy of 1 MeV. The ordinate shows
the radius $r_{\rm max}$ where we perform the matching, and
the abscissa gives the calculated phaseshift. As can clearly be seen
in the upper part of the figure,
a hard-core radius of $0.7$ fm gives the wrong phaseshifts.
To be on the safe side, we have always used a radius of $0.3$ fm in our
calculations.
Strengthened by the apparent robustness of our method, even close to
threshold, we now study the phaseshifts in the $^1S_0$ channel as
a function of the center-of-mass kinetic energy (the zero of energy
is the $NN$ threshold). \begin{figure}
\epsfysize=8cm
\centerline{\epsffile{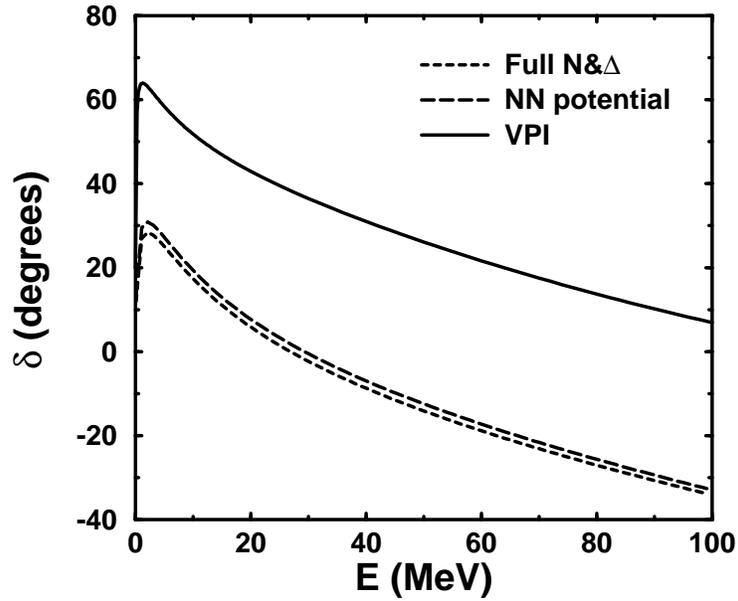}}
\caption{Phaseshifts in the $^1S_0$ channel as a function of center of mass
energy. We compare the results with and without explicit $\Delta$'s to
the phenomenological (VPI) phaseshifts.
\label{fig:1S0}}
\end{figure}
In fig.~\ref{fig:1S0} we show the phaseshifts
obtained from the full $N\Delta$ coupled channels problem  and compare
it to the phaseshifts for the adiabatic $NN$ potential.
It is important to note how close
these results are. This implies that the adiabatic potential is
a good approximation to the underlying more complicated problem.
For good measure we have also added the Arndt {\em et al\/ }phenomenological
phaseshifts \cite{ArndtRoper87,ArndtRoper87a} obtained from
the SAID database for $np$ scattering in the same figure.  Clearly
we do not have enough attraction in this channel, even though the
trend is surprisingly good. Actually we can obtain enough attraction
by increasing all parameters in the potential function $v_i$ in
Eq.~(\ref{eq:vfit}), except $b_4$, by  only 6 \%! (Keeping $b_4$ fixed
means that we do not strengthen the most repulsive part of
the interaction beyond its current value; if we increase this parameter
we find it hard to obtain agreement.)
\begin{figure}
\epsfysize=8cm
\centerline{\epsffile{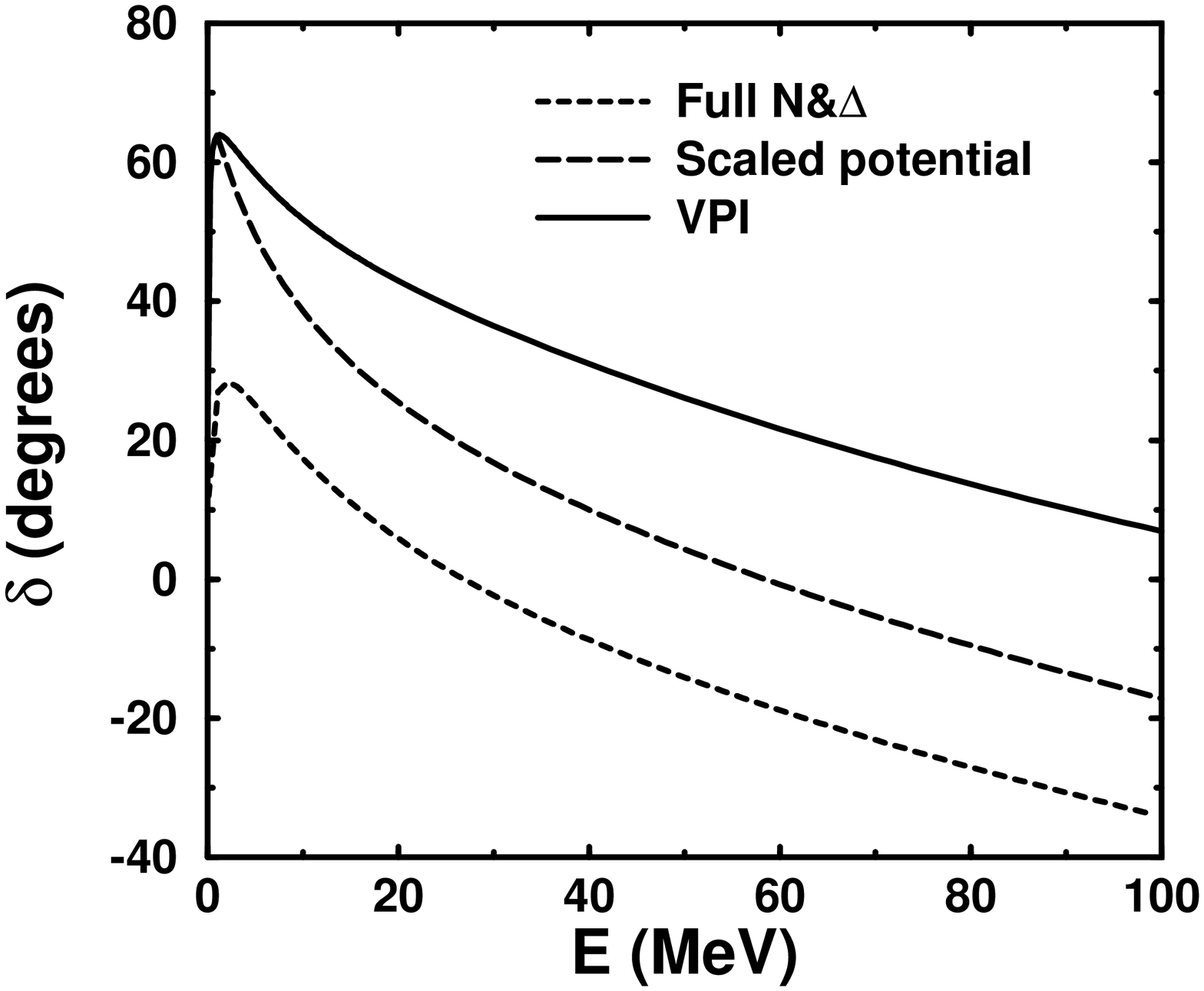}}
\caption{Phaseshifts in the $^1S_0$ channel after rescaling the potential
as discussed in the text. we compare with the phenomenological phaseshifts
and the unscaled results.
\label{fig:1S0_s}}
\end{figure}
As can be seen from Fig.~\ref{fig:1S0_s}, this gives enough attraction
near threshold. The remaining difference between our model potential
and the phenomenological phaseshifts is now mainly due to the
description of the repulsive core. This is a part of the Skyrme model
that we have not yet analyzed.

\begin{figure}
\epsfysize=14cm
\centerline{\epsffile{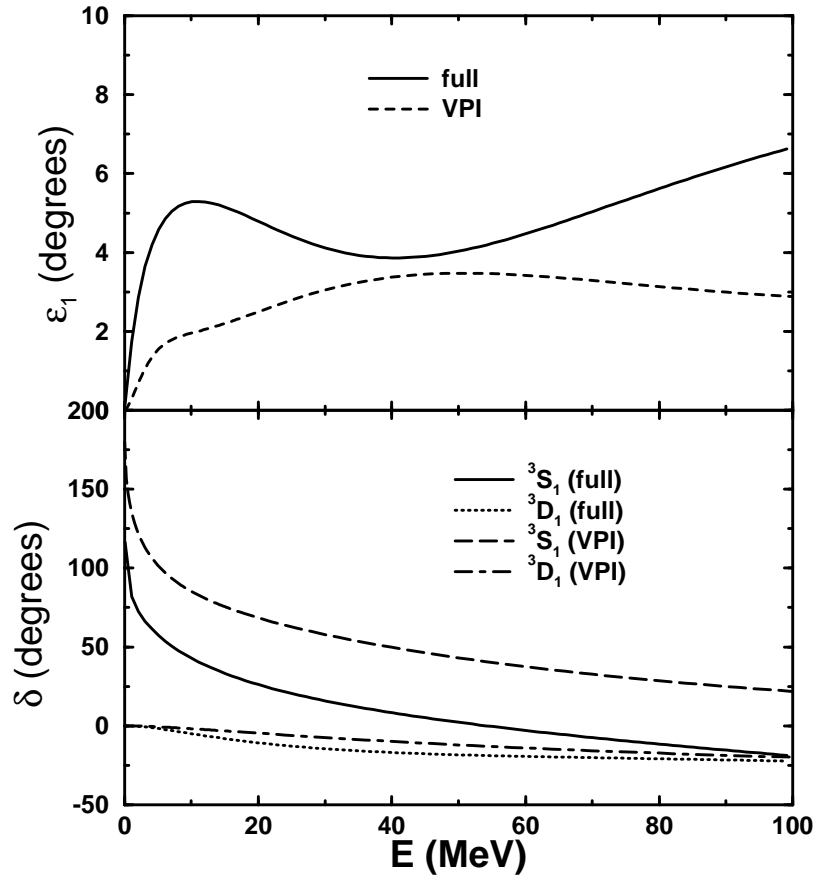}}
\caption{The mixing parameter $\epsilon_1$ and the phaseshifts
in the deuteron ($^3S_1-$$^3D_1$) channel as a function of
center-of-mass energy. We compare the phase shifts of the coupled
$N\Delta$ problem (labeled full) with the phenomenological
phase shifts of the VPI group.\label{fig:3S1}}
\end{figure}

Of course we can  perform the same analysis for the $J=1, T=1$
channel. We analyze the two-by-two $S$ matrix in terms of the standard
Stapp parameters $\delta_\pm$ and $\epsilon$ as \cite{ArndtRoper82}
\begin{equation}
S = \left(\begin{array}{ll}
\cos 2 \epsilon\; e^{2i\delta_-} &
i\sin 2 \epsilon\; e^{i(\delta_++\delta_-)} \\
i\sin 2 \epsilon\; e^{i(\delta_++\delta_-)} &
\;\;\cos 2 \epsilon\; e^{2i\delta_+}
\end{array}
\right).
\end{equation}
Again the results from the full potential and the adiabatic approximation
are very close. So close that we have not plotted them in fig.~\ref{fig:3S1}.
As can be seen the results compare favorably with the phenomenological
analysis, even though we have less attraction. To our
surprise, however, the phase-shift goes to $180^\circ$ at threshold.
This shows that our potential supports a bound state. Further analysis
shows that this occurs very close to threshold, the state is bound
by only $0.053\rm~MeV$, reflecting the fact that the attraction in the
current channel is too weak.
Again, a simple rescaling by 6 \% of all parameters but $b_4$ gives
a binding energy much closer to experiment, $E_B = 1.77~\rm MeV$.
If we improved on the calculation of the repulsive core
(which appears to be too strong in the current calculation) one
might even expect better results, since both the phaseshifts
and the binding energy of the deuteron are obtained through
the delicate balance between attraction and repulsion.
This makes it even more surprising that we obtain results
so close to experiment.

\section{Behavior near $\Delta$ production thresholds\label{sec:Delta}}
In many of the phaseshift analyses by the VPI group
(e.g., Refs.~\cite{ArndtRoper87,ArndtRoper87a})
the $T$ matrix
exhibits a resonance very near the $\Delta$ production threshold.
As better data became available some of these poles have disappeared,
but currently a few channels still seem to exhibit resonances
\cite{ArndtRoper87a}.
The abovementioned analysis has not yet been pushed to near the
double $\Delta$ production region. This region is of some interest,
however.

Theoretically it has been argued that strong attraction
in the $J=3^+$ channel is ``inevitable'' \cite{Goldman89}.
     Recently Wang {\it et al} \cite{Goldman89a} studied this question
using an approach to
baryon interactions based on nonrelativistic quark dynamics.
Using their ``quark delocalization"
scheme and color screening, they find a $\Delta-\Delta$ potential in the
S-wave, $J^\pi=3^+$, $T=0$ channel that is attractive and $400~\rm MeV$ deep.
Our model can be used to study the same question, since we have
explicit  $\Delta$'s in our calculations.
Using the same non-relativistic Hamiltonian as used in the previous
section we can discuss the problem of the
$\Delta-\Delta$ potential.

\begin{figure}
\epsfysize=8cm
\centerline{\epsffile{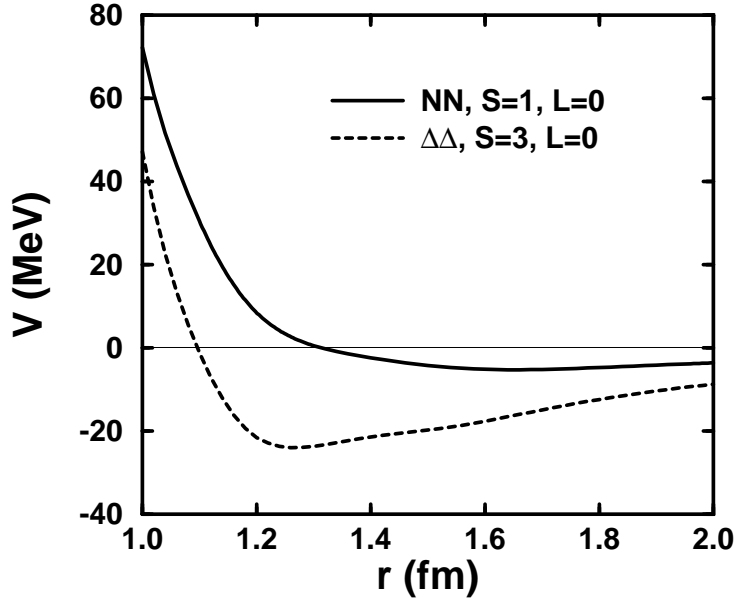}}
\caption{The $\Delta-\Delta$ potential in $J=3^+$, $L=0$ channel,
compared to the $NN$ potential in the $J=1^+$, $L=0$ channel.
Neither potential includes effects from the Born-Oppenheimer
approximation.
\label{fig:pot3p}}
\end{figure}

In Fig.~\ref{fig:pot3p} we show our result for the $\Delta-\Delta$ potential
in the $J^\pi=3^+$, $T=0$ channel as
a function of separation.
We see that there is indeed an
attractive potential in this channel, but that it is
no where near the unusually large potential reported by
Wang {\it et al}.  In this case our
potential is obtained without channel mixing (since we are studying
excited states mixing will actually decrease the binding).
If we perform a coupled channels
bound-state  calculation in the $\Delta-\Delta$ subspace only
we find a bound state with a binding energy of about $5~\rm MeV$.

We have solved for the
nucleon-nucleon phaseshifts below the $\Delta-\Delta$ threshold in the $T=0$,
$J^{\pi}=3^+$ channel, where the only two open channels are the
$NN$ $D$ and $G$ waves. Our results are only qualitative, since we use
a non-relativistic coupled-channel analysis of our potential, including the
closed channels. (Note that Wang {\em et al} also used a
non-relativistic model.)
\begin{figure}
\epsfysize=8cm
\centerline{\epsffile{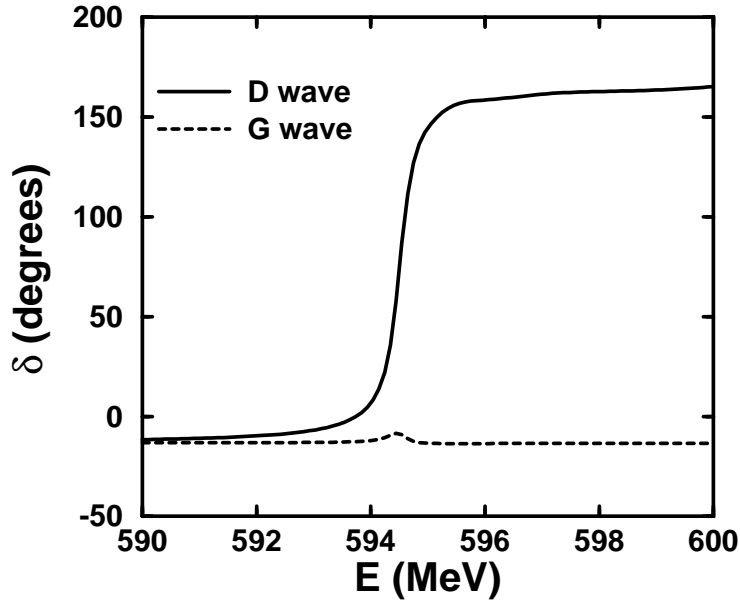}}
\caption{Phaseshifts in the $NN$ $J=3^+$ channels just below
the $\Delta-\Delta$ threshold (at 600 MeV in our calculation).
\label{fig:ps3p}}
\end{figure}
As can be seen in Fig.~\ref{fig:ps3p} we  find a sharp resonance
in the $D$-wave, $5~\rm MeV$ below the $\Delta-\Delta$ threshold.
In order to see how robust such a resonance is we include the width
of the $\Delta$ in our calculations. For a resonance close to threshold
one can make the approximation that the
 threshold energy is complex, with an imaginary part corresponding
to the width of the $\Delta$. Using this approximation we find that the
 resonance is extremely
sensitive to such a change. Even a width of $1$ MeV is enough to destroy
the resonant behavior.

How relevant are these results? We find very reasonable (if somewhat weak)
attraction in our $NN$ calculation. It is thus likely that
the relatively small attraction in the $\Delta-\Delta$ channel
is not too far from phenomenology. This would rule out
an observable non-strange dibaryon once we take into account
the width of the $\Delta$.

\section{Conclusions \label{sec:conc}}
In this paper we have shown that the Skyrme model can be used to
obtain a $NN$ interaction that is reasonable in terms of the phase
shifts it produces. The calculations reported in this paper
give phaseshifts that are surprisingly close to phenomenological
ones. With minor modifications we can obtain sufficient attraction
to get both the deuteron binding energy and the attraction in
the $^1S_0$ channel close to the experimental value. The weakest point
of the potential is actually both the description and the size
of the repulsive core. This should not surprise us since
there is a great ambiguity in the definition of the radial coordinate
at small distances, which leads to a great ambiguity in the
size of the repulsive potentials. More importantly the Skyrme
model itself should not be trusted at such distances. This
last problem can probably be solved by adding more vector mesons
to the model (the minimal extension is probably the inclusion of
the $\rho, \omega$ and $A_1$). The first problem  may
prove harder to solve, even though we have recently made some progress
on this problem using the Atiyah-Manton Ansatz in combination with
techniques of large-amplitude collective motion.
This same approach may be used to shed some light on the change of
inertial parameters with distance and orientation, which leads
to a modification of the kinetic energy part of the Hamiltonian
\cite{WaletRPA93}.
In order to understand the behavior of the inertial parameters
it would be useful to understand their behavior in the product
Ansatz first. This may lead to clues about which terms are important,
and which are not.
A comprehensive study of these parameters in the product Ansatz is currently
in progress \cite{ShaoWA93}.

\section*{Acknowledgement}
The author has benefited from many discussions with Dr. R. D. Amado.
This work was supported by the U.S. National Science Foundation.
\appendix

\section{Calculation of phaseshifts}
To calculate the phaseshifts we express the Hamiltonian in
a coupled channels matrix form,
(after re-expressing the equations in terms of $u_i=r\psi_i$)
\begin{equation}
H_{ij} = \delta_{ij}\left(\frac{-\hbar^2}{2m_i} \frac{\partial^2}{\partial r^2}
+\frac{\hbar^2L_i(L_i+1)}{2m_ir^2}+E^{\rm thres}_i\right) + V_{ij}(r).
\end{equation}
Here we require that  $V_{ij}(\infty)=0$, and thus the threshold for
channel $i$ is given by $E^{\rm thres}_i$. The integer $L_i$
is the orbital angular momentum of the given channel
In order to solve the scattering problem, we impose hard-core boundary
conditions at a small radius $r_{\rm min}$. We also assume that
the potentials $V_{ij}$ are negligibly small for a radius $r_{\rm max}$,
so that we can approximate solutions by free solutions for
$r$ larger than this value.

We integrate outwards from  $r_{\rm min}$, given the derivative at this
point. For convenience we  introduce $ v _i(r) =\partial_r  u _i(r)$.
The differential equation then reads
\begin{eqnarray}
\partial_r  u _i(r)&=&  v _i(r),  \nonumber\\
\partial_r  v _i(r)&=& -2m_i\left(\left[E-E^{\rm thres}_i-
\frac{L_i(L_i+1)}{2m_ir^2}\right] u _i(r)-V_{ij} u _j(r)\right).
\end{eqnarray}
We solve this by an explicit method (i.e., one that only uses function
values at previous values of $r$ to determine what happens at the next value).
If we keep the initial
derivative $a=\vec  v (r_{\rm min})$ as an explicit parameter, this
can be written in matrix form
\begin{equation}
\left(\begin{array}{c} \vec  u  \\ \vec  v  \end{array}\right)_{r=r_{\rm max}}
=\left(\begin{array}{cc} M^{ u  u } &  M^{ u  v } \\
                         M^{ v  u } &  M^{ v  v }\end{array}\right)
\left(\begin{array}{c} \vec 0 \\ \vec a \end{array}\right),
\end{equation}
where the matrix $M$ depends on the method used (we use a fourth order
Runge Kutta).
Due to the hard core boundary conditions we thus only need two
of the sub-matrices,
\begin{equation}
\vec  u  = M^{ u  v } \vec a,\;\;\;\vec  v  = M^{ v  v } \vec a.
\end{equation}
At this point we match to the spherical Bessel functions that describe the
asymptotic behavior of the problem, i.e., the eigenvalues of a Hamiltonian
consisting only of kinetic terms
and  centrifugal forces,
\begin{eqnarray}
 u _i(r)& = & \alpha_i k_ir h^{(1)}_{L_i}(k_ir) + \beta_i k_ir
h^{(2)}_{L_i}(k_ir).
\end{eqnarray}
Here the channel momentum $k_i=\sqrt{2m_i(E-E^{\rm thres}_i)}$,
with $k_i$ in the positive half-plane if $E<E^{\rm thres i}$.
We now solve for $\alpha_i$ and $\beta_i$ by matching at $r=r_{\rm max}$,
(It is convenient to use the notation $\tilde h^{(1)}_{L_i}(k_ir) =
k_i r h^{(1)}_{L_i}(k_ir)$)
\begin{eqnarray}
\alpha_i & = & \frac{i}{2}\left[\tilde h^{(2)'}_{L_i}(k_ir_{\rm max})
k_iM^{u v} -
\tilde h^{(2)}_{L_i}(k_ir_{\rm max}) k_iM^{v v}\right] {\bf a}
\nonumber\\
& \equiv & M_+ {\bf a}, \nonumber\\
\beta_i & = & \frac{i}{2}\left[-\tilde h^{(1)'}_{L_i}(k_ir_{\rm max})
k_iM^{u v} -
\tilde h^{(1)}_{L_i}(k_ir_{\rm max}) k_iM^{v v}\right] {\bf a}
\nonumber\\
& \equiv & M_- {\bf a}.
\end{eqnarray}
We have to impose the condition that there are no outgoing waves (which
blow up exponentially) in the closed channels, i.e., for those values
of $i$ where $k_i$ is complex. Let us label the closed channels by $c$ and
the remaining open channels by $o$. The condition that $\beta_i$ is 0 in
the closed channels can easily be implemented:
\begin{equation}
\left(\begin{array}{c}
      \beta^o \\ 0
\end{array} \right) =
\left(\begin{array}{cc}
	M_-^{oo} & M_-^{oc} \\
	M_-^{co} & M_-^{cc}
\end{array}\right) {\bf a}
\end{equation}
can be shown to imply, by inversion of the matrix, that
\begin{equation}
{\bf a} =
\left(\begin{array}{r}
	(M_-^{oo}-M^{oc}_-\frac{1}{M_-^{cc}}M^{co}_-)^{-1} \beta^o\\
	-\frac{1}{M_-^{cc}}M_-^{co}
	(M_-^{oo}-M^{oc}_-\frac{1}{M_-^{cc}}M^{co}_-)^{-1} \beta^o
\end{array}\right)
\end{equation}
We can thus relate $\alpha^o$ to $\beta^o$,
\begin{eqnarray}
 \alpha^o_i&  = & S_{ij} \beta^o_j \nonumber\\
&=&
 \left(
 \left[M_+^{oo} - M_+^{oc} \frac{1}{M_-^{cc}} M_-^{co}
 \right]
 \left[M_-^{oo} - M_-^{oc} \frac{1}{M_-^{cc}} M_-^{co}
 \right]^{-1}\right)_{ij}
 \beta_j.
 \end{eqnarray}
Here we have used the fact that our choice of phases leads directly to the
$S$ matrix.


\begin{thebibliography}{10}

\bibitem{WaletAH92}
N.~R. Walet, R.~D. Amado, and A. Hosaka, Phys. Rev. Lett. {\bf 68},  3849
  (1992).

\bibitem{WaletAmado93}
N.~R. Walet and R.~D. Amado, Phys. Rev. C {\bf 47},  498  (1993).

\bibitem{OkaHosaka92}
M. Oka and A. Hosaka, Annu.~Rev.~Nucl.~Part.~Sci. {\bf 42},  333  (1992).

\bibitem{WalhoutWambach92}
T.~S. Walhout and J. Wambach, J.~Mod.~Phys. {\bf E1},  665  (1993).

\bibitem{NymanRiska90}
E. Nyman and D. Riska, Rept.~Prog.~Phys. {\bf 53},  1137  (1990).

\bibitem{Skyrme62}
T.~H.~R. Skyrme, Nucl. Phys. {\bf 31},  556  (1962).

\bibitem{AtiyahManton93}
M.~F. Atiyah and N.~S. Manton, preprint DAMTP-92-32 (unpublished).

\bibitem{KopeliovicStern87}
V.~B. Kopeliovic and B.~E. Stern, JETP Lett. {\bf 45},  203  (1987).

\bibitem{Verbaarschot87}
J.~J.~M. Verbaarschot, Phys. Lett. B {\bf 195},  235  (1987).

\bibitem{Manton88a}
N.~S. Manton, Phys. Lett. B {\bf 192},  177  (1988).

\bibitem{BraattenCarson88}
E. Braatten and L. Carson, Phys. Rev. D {\bf 38},  3525  (1988).

\bibitem{VWWW87}
J.~J.~M. Verbaarschot, T.~S. Walhout, J. Wambach, and H.~W. Wyld, Nucl. Phys.
  {\bf A468},  520  (1987).

\bibitem{WaletRPA93}
N.~R. Walet (unpublished).

\bibitem{AmadoBO86}
R.~D. Amado, R. Bijker, and M. Oka, Phys. Rev. Lett. {\bf 58},  654  (1986).

\bibitem{OBBA87}
M. Oka, R. Bijker, A. Bulgac, and R.~D. Amado, Phys. Rev. C {\bf 36},  1727
  (1987).

\bibitem{Reidpot}
R.~V. {Reid Jr.}, Ann. Phys. {\bf 50},  411  (1968).

\bibitem{ZahedBrown86}
I. Zahed and G.~E. Brown, Phys. Rep. {\bf 142},  1  (1986).

\bibitem{Liu87} See the papers
 in {\em Chiral solitons}, edited by K.~F. Liu (World Scientific, Singapore,
  1987).

\bibitem{WalhoutWambach91}
T.~S. Walhout and J. Wambach, Phys. Rev. Lett. {\bf 67},  314  (1991).

\bibitem{HosakaOA91}
A. Hosaka, M. Oka, and R.~D. Amado, Nucl. Phys. {\bf A530},  507  (1991).

\bibitem{ShaoWA93}
B. Shao, N.~R. Walet, and R.~D. Amado (unpublished).

\bibitem{ArndtRoper87}
R.~A. Arndt, J.~S. {Hyslop III}, and L.~D. Roper, Phys. Rev. D {\bf 35},  128
  (1987).

\bibitem{ArndtRoper87a}
R.~A. Arndt, L.~D. Roper, R.~L. Workman, and M.~W. McNaughton, Phys. Rev. D
  {\bf 45},  3995  (1992).

\bibitem{ArndtRoper82}
R.~A. Arndt and L.~D. Roper, Phys. Rev. D {\bf 25},  2011  (1982).

\bibitem{Goldman89}
T. Goldman {\it et~al.}, Phys. Rev. C {\bf 39},  1889  (1989).

\bibitem{Goldman89a}
F. Wang, G.-H. Wu, L.-J. Teng, and T. Goldman, Phys. Rev. Lett. {\bf 69},  2901
   (1992).

\end{thebibliography}
\end{document}